\documentclass[pra,showpacs,print,twocolumn]{revtex4}
\usepackage{graphicx}
\usepackage{dcolumn}
\usepackage{bm}
\usepackage{txfonts}

\begin{document}
\title{Luttinger liquid coupled to Bose-Einstein condensation reservoirs}

\author{Fang Cheng$^{1}$}

\author{Guobao Zhu$^{2}$}

\author{W. M. Liu$^{2}$}

\author{Guanghui Zhou$^{1,3}$}

\email{ghzhou@hunnu.edu.cn} \affiliation{$^1$Department of Physics,
Hunan Normal University,
             Changsha 410081, China}

\affiliation{$^2$Beijing National Laboratory for Condensed Matter
Physics, Institute of Physics, Chinese Academy of Sciences, Beijing
100080, China}

\affiliation{$^3$International Center for Materials Physics,
             Chinese Academy of Sciences, Shenyang 110015, China}

\date{\today}

\begin{abstract}
We investigate the transport properties for a Luttinger liquid
coupled to two identical Bose-Einstein condensation reservoirs.
Using the approach of equation of motion for the Green function of
the system, we find that the distance between the two resonant
transmission probability peaks of the system is determined by the
bosonic interaction strengths, and the sharpness of these resonant
peaks is mainly determined by the Rabi frequency and phase of the
Bose-Einstein condensation reservoir. These results for the proposed
system involving a Luttinger liquid may build a bridge between the
controling transport properties of cold atom in atom physics and the
interacting boson transport in low-dimensional condensed matter
physics.

\end{abstract}

 \vspace{0.2cm}

\maketitle

\section{introduction}
The physics of ultracold one-dimensional (1D) Bose system is very
different from that of ordinary three-dimensional cold gases
\cite{1,2}. The possibility of Bose-Einstein condensation (BEC) in
one dimension has been discussed for the noninteracting Bose gas
\cite{3}. The interaction between bosons plays an essential role due
to the strong constraint in phase space in 1D case. Monien {\it et
al.} \cite{4} have shown that a trapped quasi-one-dimensional system
of interacting bosons under the experimental conditions can be
described by a Luttinger liquid (LL) Hamiltonian. As is known that
the low-energy physics of 1D single channel conductors can not be
described by the Fermi liquid theory if the particle-particle
interactions are taken into account \cite{5}. Such system falls into
the so-called LL regime. However, unlike the Fermi liquids, the LL
liquids may also include 1D interacting bosonic systems. Bosonic
systems can display fermion-like properties and vice versa
\cite{6,7,8}. One well-known example in the field of cold atoms is
the behavior of the Tonks-Girardeau gas \cite{9}, where the bosons
interact so strongly that they effectively behave as free fermions.

In LL model theory, the main assumption is the linearization of the
free-particle dispersion relation near the eigenenergy points of the
system. Fermionic systems which are believed to be described by the
LL model include quasi-one-dimensional organic metals \cite{10},
quantum wires \cite{11,12}, and edge states in the quantum Hall
system \cite{13}. The actual system considered is finite length and
are attached to two identical reservoirs at its end points. This is
to say, these systems are always embedded in a three-dimensional
matrix. So they will show a crossover to three-dimensional behavior
at low temperature, while the trapped 1D Bose gas would provide a
clean testing ground for the concept of LL model.

In nanoelectronics the control of electron quantum wires or quantum
dots is performed by the biased conducting leads attached to them.
But in nanobosonics the role of the ``leads" is replaced by the
finite superfluid reservoirs (given particle numbers) which can be
coupled to a particular atom by optical transitions. With regard to
this field, the dynamics of an atomic quantum dot coupled to a BEC
reservoir via laser transitions has also been studied recently
\cite{14,15}.

In this paper, we consider a system consisting of a LL coupled to
two identical BEC reservoirs. The bosons in the BEC reservoirs are
confined in a shallow trap, while the atom in the LL is confined in
a very tight potential. Atoms in both the LL and BEC reservoirs
correspond to the different internal atomic states connected by
Raman transition with Rabi frequency $\Omega$ and detuning $\delta$.
Using the approach of standard equation of motion for Green function
(GF), we investigate the frequency-dependent transport properties
for this system. Our results show that the distance between the two
resonant transmission probability peaks is determined by the
interaction strengths, and the sharpness of the resonant peak is
mainly determined by the Rabi frequency and the phase of the BEC
reservoir. The results for the proposed system involving a LL may
build a bridge between the atomic transport in atomic physics and
the interacting electron transport in low-dimensional condensed
matter physics.

\section{Model}

The total Hamiltonian of the system consists of three parts, i.e.,
\begin{equation}
H=\sum_{\alpha=L,R}H_{\alpha}+H_{cen}+H_{T},
\end{equation}
where $H_{\alpha}$ is the Hamiltonian for the isolated left or right
BEC reservoir, $H_{cen}$ is the Hamiltonian of the isolated LL, and
$H_{T}$ is the Hamiltonian describing the transfer of a particle
from the BEC to the LL.

The starting point for the calculations of $H_{\alpha} (\alpha=L,R)$
is the Hamiltonian
\begin{eqnarray}
H_{\alpha}&=&\int
dr[-\psi^{\dagger}_{\alpha}(r)\frac{1}{2m}\nabla^2\psi_{\alpha}(r)
+V(r)\psi^{\dagger}_{\alpha}(r)\psi_{\alpha}(r)]\nonumber\\
&+&\frac{1}{2}\int dr\int dr^{'}\psi^{\dagger}_{\alpha}(r)
\psi^{\dagger}_{\alpha}(r^{'})U(r-r^{'})\psi_{\alpha}(r^{'})\psi_{\alpha}(r),
\end{eqnarray}
where $\psi_{\alpha}(r)$ and $\psi^{\dagger}_{\alpha}(r)$ are
annihilation and creation operators for bosons in the BEC reservoir
respectively, $m$ is the atomic mass, $V(r)$ is the potential
confining bosons system in a trap, and $U(r-r^{'})$ is the
interaction potential between two particles in the BEC reservoir.
(We have adopted the unit of $\hbar$=1 throughout this paper.) To
take into account the quantum fluctuations of the state in which all
the atoms are condensed in a single quantum state, the operator
$\psi_{\alpha}(r)$ can be represented in the form
$\psi_{\alpha}(r)$=$\sqrt{N_{0}}\phi_{0}+\delta\psi_{\alpha}(r)$,
where $N_{0}$ is the particle number in the zero-momentum state,
$\phi_{0}$ is the wave function of the condensed state, and
$\delta\psi_{\alpha}(r)$ denotes the fluctuation operator of
momentum $k\neq 0$, i.e. the excitation above the ground state.
Within the Bogoliubov approach one assumes that
$\delta\psi_{\alpha}(r)$ is small and retains in the interaction all
terms which have two powers of $\psi_{\alpha}(r)$ or
$\psi^{\dagger}_{\alpha}(r)$. This is equivalent to including terms
which are no more than quadratic in $\delta\psi_{\alpha}(r)$ or
$\delta\psi^{\dagger}_{\alpha}(r)$.

Performing the Fourier transformation
\begin{equation}
\delta\psi_{\alpha}(r)=\sum_k a_{\alpha,k}e^{ik\cdot r},
\end{equation}
replacing $\delta\psi_{\alpha}(r)$ and
$\delta\psi^{\dagger}_{\alpha}(r)$ by $a_{\alpha,k}$ and
$a_{\alpha,k}^{\dagger}$ for $k\neq 0$, we obtain
\begin{eqnarray}
H_{\alpha}&=&\sum_{k(k\neq 0)}[(\epsilon_{\alpha,k}^{0}+U_1(k))
(a_{\alpha,k}^{\dagger}a_{\alpha,k}+a_{\alpha,-k}^{\dagger}a_{\alpha,-k})
\nonumber\\
&+&U_2(k)(a_{\alpha,k}^{\dagger}a_{\alpha,-k}^{\dagger}+a_{\alpha,k}a_{\alpha,-k})],
\end{eqnarray}
where $\epsilon_{\alpha,k}^{0}$ is the single particle energy, and
$U_1(k)$ ($U_2(k)$) is the Fourier transformation of $U(r-r^{'})$.
Here the operators $a_{\alpha,k}$ and $a_{\alpha,k}^{\dagger}$ are
destroy and create bosons in the state with momentum $k$ satisfy the
usual Bose commutation relations.

We take the effect when two atoms are close to each other into
account by using the effective interaction, and the Hartree-Fock
terms are both equal to $n_0U_0$ in which $n_{0}$ is the number
density of the BEC, $U_{0}$ the contact interaction in Hatree-Fock
approximation, so the Hamiltonian for the isolated left or right BEC
reservoir reads \cite{16}
\begin{eqnarray}
H_{\alpha}&=&\sum_{k(k\neq 0)}[(\epsilon_{\alpha,k}^{0}+n_0U_0)
(a_{\alpha,k}^{\dagger}a_{\alpha,k}+a_{\alpha,-k}^{\dagger}a_{\alpha,-k})
\nonumber\\&+&n_0U_0(a_{\alpha,k}^{\dagger}a_{\alpha,-k}^{\dagger}+
a_{\alpha,k}a_{\alpha,-k})],
\end{eqnarray}
where the single particle energy $\epsilon_{\alpha,k}^{0}=k^2/(2m)$,
and $a_{\alpha,k}^{\dagger}$ $(a_{\alpha,k})$ is the creation
(annihilation) operator of the bosons in the left or right
reservoir. Note that the prime on the sum indicates that it is to be
taken only over one half of momentum space.

Considering the Raman detuning between LL and the reservoir and the
phase of the reservoir, the Hamiltonian of the isolated left or
right BEC reservoir, i.e., the first term $H_{\alpha}$
$(\alpha=L,R)$ of Eq. (1), is given by
\begin{eqnarray}
H_{\alpha}&=&\sum_{k(k\neq 0)}[(\epsilon_{\alpha,k}+|\Delta|)
(a_{\alpha,k}^{\dagger}a_{\alpha,k}+a_{\alpha,-k}^{\dagger}a_{\alpha,-k})\nonumber\\
&+&(\Delta a_{\alpha,k}^{\dagger}a_{\alpha,-k}^{\dagger}+\Delta^{*}
a_{\alpha,k}a_{\alpha,-k})],
\end{eqnarray}
where $\epsilon_{\alpha,k}=\epsilon_{\alpha,k}^{0}+\delta_{\alpha}$
in which $\delta_{\alpha}$ is the Raman detuning between LL and the
reservoir $\alpha$; $\Delta=n_{0}U_{0}e^{i\phi_{\alpha}}$ in which
$\phi_{\alpha}$ is the phase of the reservoir $\alpha$.  In the
absence of the driver ($\delta_{L}$=$\delta_{R}$), the particles can
also pass through the system because of the phase difference between
two BEC reservoirs.

For the Bose gas in a cylindrical symmetric trap confined to the $z$
axis by a tight trapping potential in $x$-$y$ plane, if the
extension $L$ of the trap in the $z$ direction is much larger than
its radius $R$ and the temperature is much lower than the energy of
the lowest radial excitation, the ground state is described by a LL
\cite{4}. The starting point for the calculations of $H_{cen}$ is
also the Hamiltonian
\begin{eqnarray}
H_{cen}&=&\int dr[-\psi^{\dagger}_{1}(r)\frac{1}{2m}\nabla^2\psi_{1}(r)
+V^{'}(r)\psi^{\dagger}_{1}(r)\psi_{1}(r)]\nonumber\\
&+&\frac{1}{2}\int dr\int dr^{'}\psi^{\dagger}_{1}(r)
\psi^{\dagger}_{1}(r^{'}) U^{'}(r-r^{'})\psi_{1}(r^{'})\psi_{1}(r),
\end{eqnarray}
where $\psi_{1}(r)$ and $\psi^{\dagger}_{1}(r)$ are annihilation and
creation operators for bosons in the LL respectively, $V^{'}(r)$ is
the potential confining bosons system in a trap, and
$U^{'}(r-r^{'})$ is the interaction potential between two particles
in the LL. Through the same procedures as above, the operator
$\psi_{1}(r)$ can be represented in the form
$\psi_{1}(r)$=$\sqrt{N^{'}_{0}}\phi^{'}_{0}+\delta\psi_{1}(r)$. And
the fluctuation operator
\begin{equation}
\delta\psi_{1}(r)=\sum_q b_{q}e^{iq\cdot r},
\end{equation}
where the operators $b_{q}$ and $b_{q}^{\dagger}$ that destroy and
create bosons in the state with momentum $q$ satisfy the usual Bose
commutation relations
\begin{equation}
[b_{q},b^{\dagger}_{q^{'}}]=\delta_{q,q^{'}},~~
[b_{q},b_{q^{'}}]=0,~~ [b_{q}^{\dagger},b_{q^{'}}^{\dagger}]=0.
\end{equation}
Using the Bogoliubov approach and replacing $\delta\psi_{1}(r)$ and
$\delta\psi^{\dagger}_{1}(r)$ by $b_{q}$ and $b_{q}^{\dagger}$ for
$q\neq 0$, we obtain
\begin{eqnarray}
H_{cen}&=&\sum_{q(q\neq 0)}[(\epsilon_{q}^{0}+U_1(q))
(b_{q}^{\dagger}b_{q}+b_{-q}^{\dagger}b_{-q})\nonumber\\&+&U_2(q)
(b_{q}^{\dagger}b_{-q}^{\dagger}+b_{q}b_{-q})],
\end{eqnarray}
where $\epsilon_{q}^{0}$ is the single particle energy, and $U_1(q)$
($U_2(q)$) is the Fourier transformation of $U^{'}(r-r^{'})$.

It is known that in the LL model, the main assumptions are: (1) the
linearization of the dispersion relation; (2) only small momenta
exchanges included. In terms of the two assumptions, the Hamiltonian
of the isolated LL, i.e., the second term $H_{cen}$ of Eq. (1), is
given by \cite{17}
\begin{eqnarray}
H_{cen}&=&\sum_{q>0}\Big[(v_L(q+k_{L})+\frac{V_4}{2\pi})
(b_{q}^{\dagger}b_{q}+b_{-q}^{\dagger}b_{-q})\nonumber\\&+&\frac{V_2}{2\pi
}(b_{q}^{\dagger}b_{-q}^{\dagger}+b_{q}b_{-q})\Big].
\end{eqnarray}
Here the single particle energy $\epsilon_{q}^{0}$=$v_L(q+k_{L})$
because of the linearization of the dispersion relation. In
Hamiltonian (11), $b_{q}^{\dagger}$ ($b_{q}$) is the creation
(annihilation) operator of the bosons in the LL, $v_{L}$ is the
eigen-velocity in the channel, $k_{L}$ is the eigen-wavevector, and
$V_{j}$ ($j=2,4$) is the interaction potential when $q\sim0$ in
which $V_{2}$ represents the interaction potential between the left-
and right-moving boson branches, while $V_{4}$ represents the
interaction potential within a momentum branch.

Note that the LL arisen in our system describes Bose system, so the
operators $b_q$ and $b^{\dagger}_q$ correspond to the destruction or
creation of an individual particle (i.e.,boson). However, when the
LL describes the interacting electrons in one dimension, an
individual particle is a fermion. By means of the bosonization
technique, we can also write the Hamiltonian in terms of boson
operators. But the operators $b_q$ and $b^{\dagger}_q$ are linear
combination of the density fluctuations $\rho_q = \sum\limits_k
c^\dagger_{k} c_{k+q}$, so they conserve the number of fermion
particle and do not correspond to the destruction or creation of an
individual particle.

The Bose field can also be describe by its density-phase
representation: $\psi_{B}(r)$=$\sqrt{\rho(r)}e^{i\theta(r)}$.
Expanding in small fluctuations of the phase $\delta\theta$ and the
density $\delta\rho$ around the saddle-point solution,
$\psi_{B}(r)$=$\sqrt{\rho_0+\delta\rho(r)}e^{i[\theta_0+\delta\theta(r)]}$.
The density fluctuation operator $\delta\rho$ and the phase
fluctuation operator $\delta\theta$ form a pair of conjugate
operators [$\delta\theta(z)$,
$\delta\rho(z^{'})$]=$i\delta(z-z^{'})$. With the same approximation
as the equations of motion in Ref [4], we can also express the
Hamiltonian of the isolated LL as (Eq. (10) in Ref. [4])
\begin{equation}
H_{cen}=\int dz\bigg[\frac{\rho}{2m}(\partial_z\delta\theta)
+\frac{\kappa}{2\rho^2}\delta\rho^2\bigg],
\end{equation}
where $\rho$ is the number of particles per unit length, $m$ is the
atomic mass, and $\kappa$ is the compressibility.

The starting point for the calculations of $H_{T}$ which describes
the transfer of a particle from the BEC to the LL, is the
Hamiltonian
\begin{eqnarray}
H_{T}&=&\Omega_{L}\psi^{\dagger}_{L}(r_1)\psi_{1}(r_1)
+\Omega^{*}_{R}\psi^{\dagger}_{1}(r_2)\psi_{R}(r_2)\nonumber\\
&+&\Omega^{*}_{L}\psi_{1}^{\dagger}(r_1)\psi_{L}(r_1)
+\Omega_{R}\psi_{R}^{\dagger}(r_2)\psi_{1}(r_2),
\end{eqnarray}
where $\Omega_{\alpha} (\alpha=L,R)$ is Rabi frequency. Here we have
assumed that the atom in the LL is coupled to atoms in the reservoir
$\alpha$ via Raman transition with Rabi frequency. Using the
Bogoliubov approach and replacing $\delta\psi_{\alpha}(r)$ and
$\delta\psi^{\dagger}_{\alpha}(r)$ by $a_{\alpha,k}$ and
$a_{\alpha,k}^{\dagger}$ for $k\neq 0$, we obtain
\begin{eqnarray}
H_{T}&=&\Omega_{L}\sum_ke^{-ik\cdot
r_1}a^{\dagger}_{L,k}\delta\psi_{1}(r_1)
\nonumber\\&+&\Omega^{*}_{R}\delta\psi^{\dagger}_{1}(r_2)\sum_k
a_{R,k}e^{ik\cdot r_2}+h.c.,
\end{eqnarray}
where h.c. denotes the Hermitian conjugate. In terms of the operator
$\psi_{1}(r)$=$\sqrt{N^{'}_{0}}\phi^{'}_{0}+\delta\psi_{1}(r)$
=$\sqrt{\rho_0+\delta\rho(r,t)}e^{i[\theta_0+\delta\theta(r,t)]}$,
if replacing $\delta\psi_{1}(r)$ by $e^{i\delta\theta}(r)$, the
Hamiltonian (14) becomes
\begin{eqnarray}
H_{T}&=&\Omega_{L}\sum_ke^{-ik\cdot
r_1}a^{\dagger}_{L,k}e^{i\delta\theta(r_1)}
\nonumber\\&+&\Omega^{*}_{R}e^{-i\delta\theta(r_2)}\sum_k
a_{R,k}e^{ik\cdot r_2}+h.c..
\end{eqnarray}
And replacing $\delta\psi_{1}(r)$ and $\delta\psi^{\dagger}_{1}(r)$
by $b_{q}$ and $b_{q}^{\dagger}$ for $q\neq 0$, the Hamiltonian (14)
becomes
\begin{equation}
H_{T}=\sum_{k,\alpha,q}(\Omega_{\alpha}a_{\alpha,k}^{\dagger}b_{q}
+\Omega_{\alpha}^{*}b_{q}^{\dagger}a_{\alpha,k}).
\end{equation}
Because the operator $a^{\dagger}_{\alpha,k}$ correspond to a
creation of a particle in the BEC reservoir and the operator $b_q$
correspond to a destruction of a particle in the LL, a terms such as
$b_q^\dagger a_{\alpha,k}$ would thus correspond to a destruction of
a particle in the BEC and a creation of particles in the LL.

However, for a fermionic LL, if there is transfer of a particle from
the BEC reservoirs to the fermionic LL, it must have a different
form,
\begin{eqnarray}
H_{T}&=&\Omega_{L}\psi^{\dagger}_{L}(r_1)\psi_{F}(r_1)
+\Omega^{*}_{R}\psi^{\dagger}_{F}(r_2)\psi_{R}(r_2)\nonumber\\
&+&\Omega^{*}_{L}\psi_{F}^{\dagger}(r_1)\psi_{L}(r_1)
+\Omega_{R}\psi_{R}^{\dagger}(r_2)\psi_{F}(r_2),
\end{eqnarray}
where $\psi_{F}$=$\sum\limits_q c_{q} e^{iq\cdot r}$ is the Fermi
annihilation operator. Using the bosonization technique, the Fermi
annihilation operator can be written as \cite{18}
\begin{equation}
\Psi_{F}(r)\sim \sum_{p=\pm1}e^{ip\vartheta(r)+i\phi(r)}\sim
exp(i\sum_q(e^{iq\cdot r}b_q+h.c.)),
\end{equation}
where $\vartheta(r)$ and $\phi(r)$ obey the commutation relations
$[\phi(r),\phi(r^{'})]$=$[\vartheta(r),\vartheta(r^{'})]$=0 and
$[\phi(r), \vartheta(r^{'})]$=$isgn(r-r^{'})/2$.

Note that the total Hamiltonian $H$ in our system is equal to the
sum of Eq. (6), Eq. (11) and Eq. (16).

\section{Formulation}

In terms of the Heisenberg equation of motion, the current of the
reservoir $\alpha$ can be written as
\begin{eqnarray}
I_{\alpha,k}(t)&=&\langle\frac{dN_{\alpha,k}}{dt}\rangle
=-i\langle[N_{\alpha,k},H]\rangle\nonumber\\
&=&-i\langle\sum_{kq}(\Omega_{\alpha}a_{\alpha,k}^{\dagger}b_{q})+h.c.\rangle,
\end{eqnarray}
where
$N_{\alpha,k}=\sum\limits_{k}a_{\alpha,k}^{\dagger}a_{\alpha,k}$ is
the total number operator for the boson in the reservoir $\alpha$.
Defining a 2$\times$2 GF $G_{q,k\alpha}^{<}(t)$, then the current of
the reservoir $\alpha$ becomes
\begin{equation}
I_{\alpha,k}(t)=-i\sum_{kq}(\Omega_{\alpha}G_{q,k\alpha}^{<}(t))_{11}+h.c.,
\end{equation}
where $(G_{q,k\alpha}^{<}(t))_{11}=\langle
a_{\alpha,k}^{\dagger}b_{q}\rangle$ is the element in the first row
and the first column of GF $G_{q,k\alpha}^{<}(t)$.

Similarly,
\begin{equation}
I_{\alpha,-k}(t)=-i\sum\limits_{kq}(\Omega_{\alpha}G_{q,k\alpha}^{<}(t))_{22}+h.c.,
\end{equation}
where $(G_{q,k\alpha}^{<}(t))_{22}$ is the element in the second row
and the second column of GF $G_{q,k\alpha}^{<}(t)$. Since the
current is conserved, the currents of the bosons with momentum $k$
and $-k$ are equal, i.e., $I_{\alpha,k}=I_{\alpha,-k}$.

Using the theorem of analytic continuation, we have
\begin{eqnarray}
G_{q,k\alpha}^{<}(t,t^{'})&=&G^{r}(t,t_{1})\Omega_{\alpha}^{*}g_{\alpha,k}^{<}
(t_{1},t^{'})\nonumber\\
&+&G^{<}(t,t_{1})\Omega_{\alpha}^{*}g_{\alpha,k}^{a}(t_{1},t^{'}),
\end{eqnarray}
where $G^{</r}$ is 2$\times$2 lessor/retarded GF of LL with coupling
between the LL and the reservoir, while $g_{\alpha,k}^{<,a}$ is
2$\times$2 lessor/advanced GF of the isolated BEC reservoir,
respectively. The GF $g_{\alpha ,k}^{<}(E)$ in Fourier space is
given by
\begin{equation}
g_{\alpha ,k}^{<}(E)=[g_{\alpha ,k}^{a}(E)-g_{\alpha
,k}^{r}(E)]f_{\alpha}(E),
\end{equation}
where $f_{\alpha}(E)$ ($\alpha=L,R$) is the Bose distribution
function. And based on the Landauer-Buttiker formula
\cite{19,20,21}, the current in the Fourier space can be written as
\begin{equation}
I_{\alpha}=-\int_{\delta_{m}}^{\infty}\frac{dE}{2\pi}[f_{L}(E)-f_{R}(E)]T(E),
\end{equation}
where $E$ is the energy of the incident boson from the reservoir
$\alpha$. And with the help of Keldysh equation:
$G^{<}=G^{r}\Sigma^{<}G^{a}$, the transmission probability $T(E)$ is
solved as
\begin{equation}
T(E)=4Tr[\Sigma_{L}^{r}(E-\delta_{L})\sum\limits_{q}G^{r}(E)
\Sigma_{R}^{r}(E-\delta_{R})\sum\limits_{q}G^{a}(E)].
\end{equation}
Here $\Sigma^{</r}$ is lessor/retarded self energy, respectively.
The retarded self energy is given by
$\Sigma^{r}=\sum\limits_{\alpha}\Sigma_{\alpha}^{r}$ where
$\Sigma_{\alpha}^{r}$ is retarded self energy of reservoir $\alpha$
and defined as
\begin{equation}
\Sigma_{\alpha}^{r}=\sum\limits_{k}t_{k,\alpha}^{*}g_{\alpha,k}^{r}t_{k,\alpha},
\end{equation}
where \begin{equation} t_{k,\alpha}=\left(\begin{array}{cccc}
\Omega_{\alpha}  &   0\\
0  &   -\Omega^{*}_{\alpha}\\
\end{array}\right),
\end{equation}
and $g_{\alpha,k}^{r}$ is the retarded GF of the isolated left or
right BEC reservoir and defined as
\begin{eqnarray}
g_{\alpha,k}^{r}(t,t^{'})&=&-i\theta(t-t^{'})\nonumber\\&\times&
\left(\!\!\begin{array}{cccc}
\langle\{ a_{\alpha,k}(t),a_{\alpha,k}^{\dagger}(t^{'})\} \rangle   &   \langle\{ a_{\alpha,k}(t),a_{\alpha,-k}(t^{'})\} \rangle\\
\langle\{ a_{\alpha,-k}^{\dagger}(t),a_{\alpha,k}^{\dagger}(t^{'})\} \rangle  &   \langle\{ a_{\alpha,-k}^{\dagger}(t),a_{\alpha,-k}(t^{'})\} \rangle\\
\end{array}\!\!\right).
\end{eqnarray}
In terms of the equation of motion for the GF, $g_{\alpha,k}^{r}$ in
the Fourier space can be written as
\begin{equation}
g_{\alpha,k}^{r}= \frac{1}{E^2-\epsilon_{\alpha
k}^{2}-2|\Delta|\epsilon_{\alpha k}}\left(\begin{array}{cccc}
E+\epsilon_{\alpha
k}+|\Delta|   &   -\Delta\\
-\Delta^{*}  &   -E+\epsilon_{\alpha
k}+|\Delta|  \\
\end{array}\right).
\end{equation}

Defining $g_{\alpha}^{r}$=$\sum\limits_{k}g_{\alpha,k}^{r}$, by
straightforward calculation, we obtain
\begin{eqnarray}
g_{\alpha}^{r}&=&\rho_{\alpha}\left(
\begin{array}{cc}
E & -\Delta \\
-\Delta^{\ast} & -E%
\end{array}
\right) \int\!\frac{d(\epsilon_{\alpha,k}+\left\vert
\Delta\right\vert )}{(E+i0^{+})^{2}-(\epsilon_{\alpha,k}+\left\vert
\Delta\right\vert )^{2}+\left\vert \Delta\right\vert ^{2}
}\nonumber\\
&+&\!\!\!\rho_{\alpha}\left(
\begin{array}{cc}
\epsilon_{\alpha,k}+\left\vert
\Delta\right\vert & 0 \\
0 & \epsilon_{\alpha,k}+\left\vert \Delta\right\vert
\end{array}
\right) \int\!\!\!\frac{d(\epsilon_{\alpha,k}+\left\vert
\Delta\right\vert
)}{(E+i0^{+})^{2}\!-(\epsilon_{\alpha,k}+\left\vert
\Delta\right\vert )^{2}\!+\left\vert \Delta\right\vert ^{2}
}\nonumber\\
\end{eqnarray}
Here we have changed $\sum\limits_{k}$ into an integral $\int
d\epsilon_{\alpha, k} \rho_{\alpha}$ with the help of the density of
the states in the BEC reservoir $\rho_{\alpha}$. The second term of
Eq. (30) vanishes, because in the second term both
$g_{\alpha,11}^{r}$ and $g_{\alpha,22}^{r}$ are odd functions of
$\epsilon_{\alpha,k}+\left\vert \Delta\right\vert $. So we finally
obtain
\begin{eqnarray}
g_{\alpha}^{r}=\rho_{\alpha}\left(
\begin{array}{cc}
E & -\Delta \\
-\Delta^{\ast} & -E%
\end{array}
\right) \int \frac{d(\epsilon_{\alpha,k}+\left\vert
\Delta\right\vert )}{(E+i0^{+})^{2}-(\epsilon_{\alpha,k}+\left\vert
\Delta\right\vert )^{2}+\left\vert \Delta\right\vert ^{2} }.
\end{eqnarray}
 In the following,
calculating the intergral by using the residual theorem, Eq. (31)
can be reduced as
\begin{eqnarray}
g_{\alpha}^{r}=\rho_{\alpha}\frac{-i\nu\pi}{\sqrt{E^{2}+\left\vert
\Delta\right\vert ^{2}}}\left(
\begin{array}{cc}
E & -\Delta \\
-\Delta^{\ast} & -E%
\end{array}
\right),
\end{eqnarray}
with $\nu=1$ for $E>0$ and $\nu=-1$ otherwise.

Inserting Eqs. (27) and (32) into (26), we obtain
\begin{equation}
\Sigma_{\alpha}^{r}=-\frac{i\nu\Gamma_{\alpha}}{2}\frac{1}
{\sqrt{E^2+|\Delta|^2}}\left(\begin{array}{cccc}
E  &   \Delta\\
\Delta^{*}  &   -E\\
\end{array}\right),
\end{equation}
where the linewidth function
$\Gamma_{\alpha}$=2$\pi|\Omega_{\alpha}^{2}|\rho_{\alpha}$. Under
the so called wide-band approximation, the self energy of the lead
is not sensitive to the energy and can be taken as a constant
independent of the energy $E$. The non-diagonal term in the
expression of the self energy $\Delta$=$|\Delta|e^{i\phi}$ is
$|\Delta|$ when the phase of the BEC reservoir $\phi$=0.

In Eq. (25), $G^{r}(E)$ and $G^{a}(E)$ denote the Fourier transforms
of the GF $G^{r}(t)$ and $G^{a}(t)$ respectively. $G^{r}$ can be
obtained by Dyson equation in matrix form
\begin{equation}
G^{r}=G_{0}^{r}+G^{r}\Sigma^{r}G_{0}^{r},
\end{equation}
where $G_{0}^{r}$ is the retarded GF of the isolated LL which is
defined as
\begin{eqnarray}
G_{0}^{r}(t,t^{'})&=&-i\theta(t-t^{'})\nonumber\\&\times&
\left(\begin{array}{cccc}
\langle\{ b_{q}(t),b_{q}^{\dagger}(t^{'})\} \rangle   &
\langle\{ b_{q}(t),b_{-q}(t^{'})\} \rangle\\
\langle\{ b_{-q}^{\dagger}(t),b_{q}^{\dagger}(t^{'})\} \rangle  &
\langle\{ b_{-q}^{\dagger}(t),b_{-q}(t^{'})\} \rangle\\
\end{array}\right).
\end{eqnarray}
In terms of the equation of motion for the GF, we can obtain
\begin{widetext}
\begin{equation}
G_{0}^{r}= \frac{\displaystyle 1}{\displaystyle
E^2-(v_{L}q(\frac{1}{g^2}+1)/2+v_{L}k_{L})^2+(v_{L}q(\frac{1}{g^2}-1)/2)^2}
\left(\begin{array}{cccc}
E+v_{L}q(\frac{1}{g^2}+1)/2+v_{L}k_{L}   &   -v_{F}q(\frac{1}{g^2}-1)/2\\
-v_{L}q(\frac{1}{g^2}-1)/2 &  -E+v_{L}q(\frac{1}{g^2}+1)/2+v_{L}k_{L}  \\
\end{array}\right),
\end{equation}
\end{widetext} where the parameter $g$ is the strength of the
interaction which is defined as $g=(1+V/(\pi v_L))^{-1/2}$. Here we
have assumed that $V_{2}$=$V_{4}$=$V$. This definition follows that
of the fermions. The LL parameters $g$ also can be extracted from
the Lieb-Liniger equation \cite{22,23}. For repulsive bosons, $g$=1
corresponds to the hard-core limit, while $g$$>$1 for repulsion,
with $g$$\rightarrow$$\infty$ in the limit of weak interactions. In
the case of fermion, non-interacting fermion corresponds to $g$=1
and repulsive interaction corresponds to $g$ $<$ 1.

\section{quantum transport}
In the following we show some numerical examples calculated
according to Eq. (25) for the transport properties of this system
with the experimental parameters: \cite{24} for $^{87}$Rb, $T$=1 nK
and the eigenenergy of the LL $E_{L}$=2.0 kHz. By analyzing the form
of $G^r_{0}$ for the isolated LL, because $q/k_{L}$$\approx$0 or
$q/k_{L}$$\approx$2, and the energy is equal the sum of the
excitation energy and the eigenenergy of the LL, there should appear
peaks near the eigenenergy and near three times of the eigenenergy
in the transmission probability versus the energy of the incident
boson. And peaks near $E/E_{L}$=1 or near $E/E_{L}$=3 will evidently
differ from the resonant peak in the case of quantum dot coupled to
BEC reservoirs. Because $q$ has a range of values, this will open
some new channels for transmission. Here we will consider the
symmetric case, i. e., $\Gamma_L$=$\Gamma_R$=$\Gamma$/2.

\begin{figure}
\center
\includegraphics[width=2.5in]{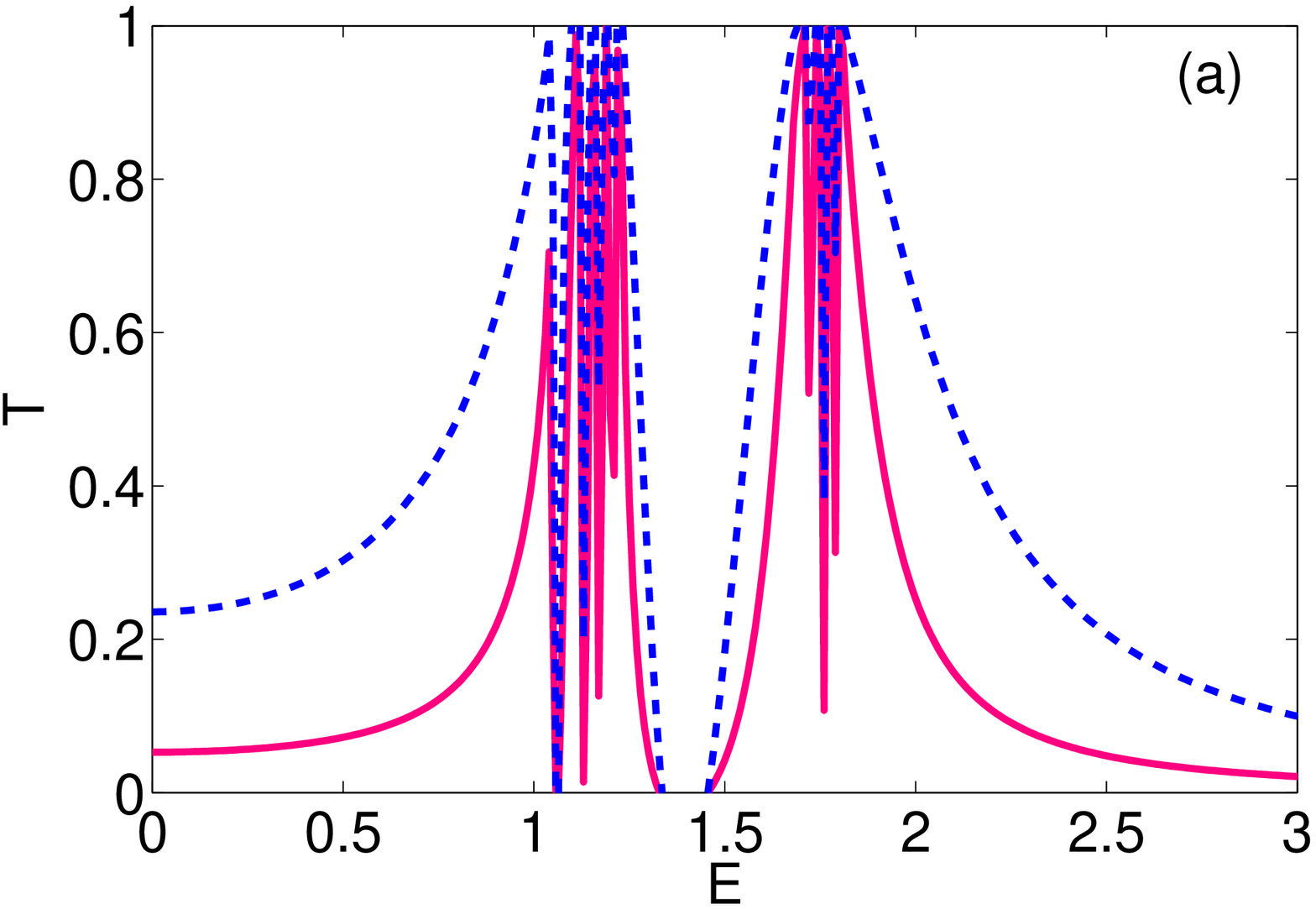}
\includegraphics[width=2.5in]{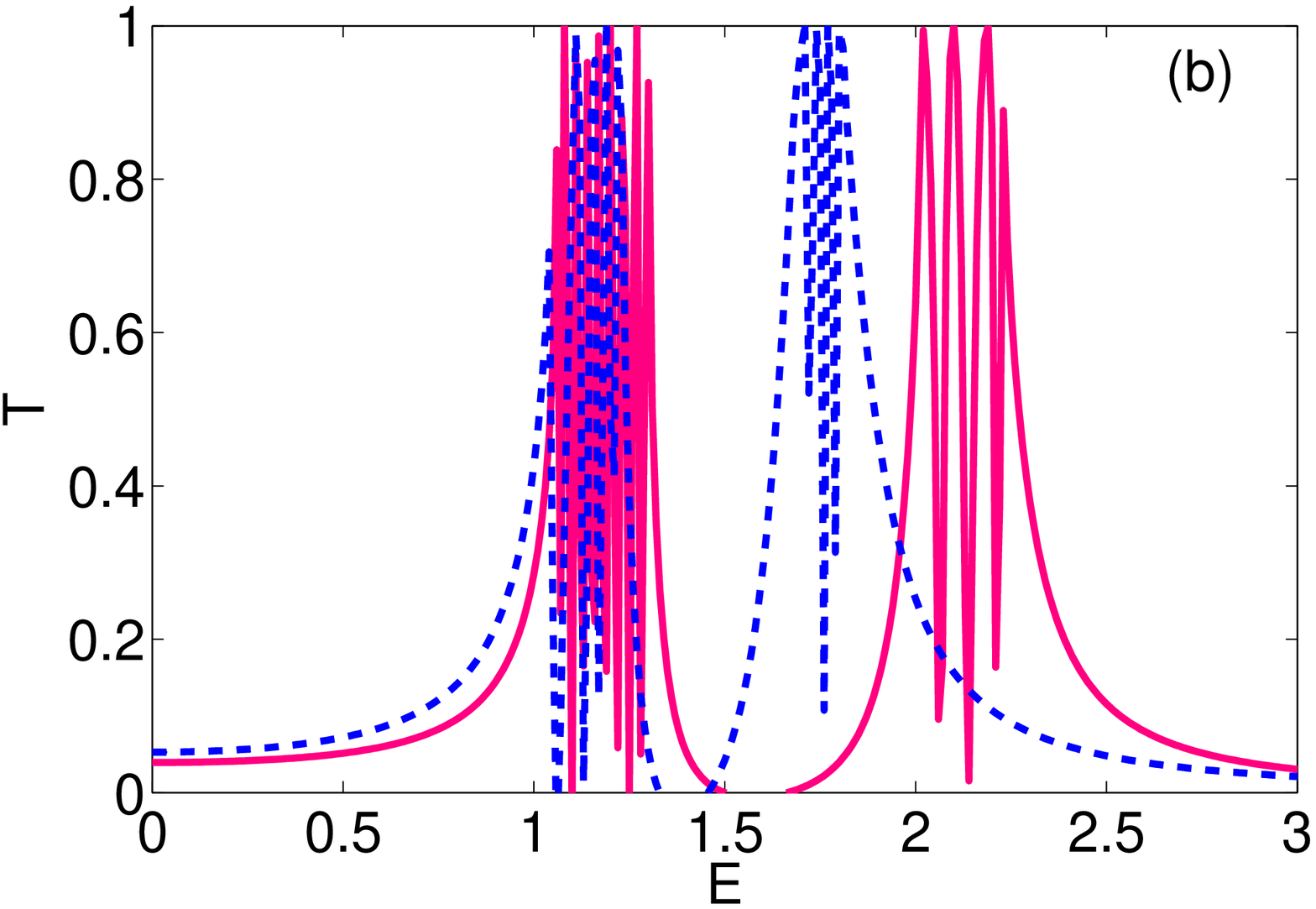}
\includegraphics[width=2.5in]{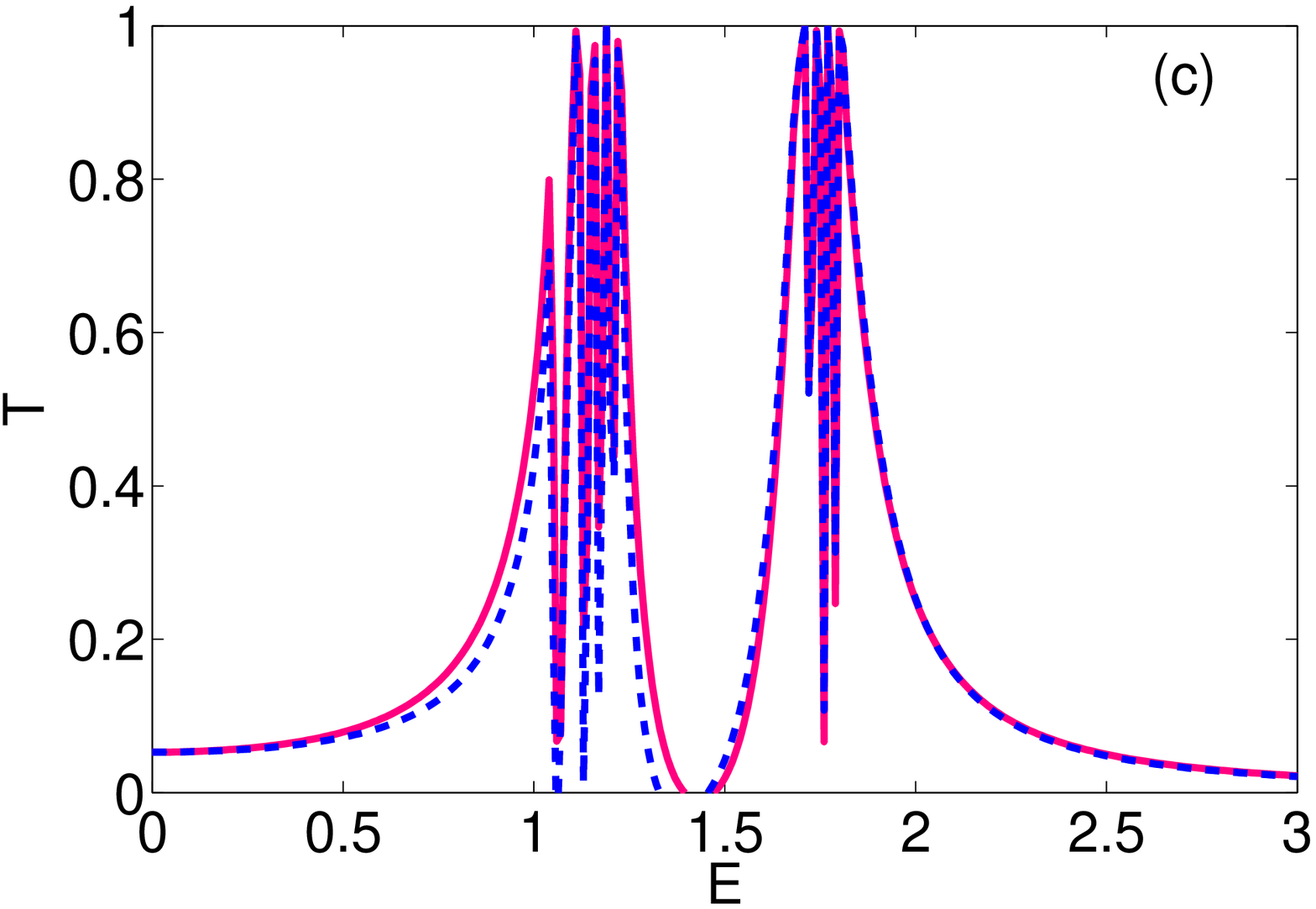}
\caption{(Color Online) The transmission probability $T$ as a
function of incident boson energy $E$ (in units of $E_{L}$,
$E_{L}$=2.0 kHz) where the parameters are $\delta_L$=$\delta_R$=0
and $\phi$=0, (a) for two different Rabi frequencies $\Omega$=0.02
kHz (red solid line) and 0.03 kHz (blue dashed line) with
$|\Delta|=2\pi\times$0.41 kHz and $g$=10, (b) for two different
interaction strengths $g$=2 (red solid line) and 10 (blue dashed
line) with $\Omega$=0.02 kHz and $|\Delta|=2\pi\times$0.41 kHz, and
(c) for two different $|\Delta|$=0.2$\pi$ kHz (red solid line) and
0.82$\pi$ kHz (blue dashed line) with $\Omega$=0.02 kHz and $g$=10,
respectively.}
\end{figure}
Fig. 1 illustrates the transmission probability T as a function of
the incident boson energy $E$ (in units of $E_{L}$) where the
parameters are $\delta_L$=$\delta_R$=0 and $\phi$=0. For the system
with fixed $|\Delta|$=2$\pi\times0.41$ kHz and $g$=10, Fig. 1(a)
shows the dependence of two different Rabi frequency $\Omega$=0.02
kHz (red solid line) and 0.03 kHz (blue dashed line) on the
transmission probability, respectively. From this figure we can find
that the resonant peak becomes wider as the Rabi frequency
increases. It is because $|\Omega|^2\propto \Gamma$, and $\Gamma$
describes how well the reservoir is in contact with the LL. The
larger linewidth function corresponds to the stronger coupled case,
and the stronger coupling corresponds to the wider the resonant
peak. Fig. 1(b) shows the result of the transmission probability $T$
versus $E$ with fixed Rabi frequency $\Omega$=0.02 kHz and
$|\Delta|$=2$\pi\times$0.41 kHz for two different interaction
strengths, where the red solid line for $g$=2 and the blue dashed
line for $g$=10, respectively. The distance between the two resonant
peaks become smaller when the particle-particle interaction
parameter $g$ is larger. From this figure, we can conclude that the
interaction parameter plays an important role on the relative
position of the two resonant peaks. Fig. 1(c) illustrates T as a
function of $E$ with fixed Rabi frequency $\Omega$=0.02 kHz and
fixed interaction strength $g$=10 for two different
$|\Delta|$=0.2$\pi$ kHz (red solid line) and 0.82$\pi$ kHz (blue
dashed line), respectively. From this figure, we can not find more
visible difference between the two cases, which show that the width
and height of the peaks are not sensitive to $|\Delta|$.

Fig. 2 shows the dependence of three different phases of BEC
reservoir on the transmission probability with fixed Rabi frequency
$\Omega$=0.02 kHz, $|\Delta|$=2$\pi \times$0.41 kHz and $g$=10,
where the blue line for $\phi$=0 or $\phi$=2$\pi$, the red dashed
line for $\phi$=$\pi$/2 or $\phi$=3$\pi$/2 and the green dashed line
for $\phi$=$\pi$, respectively. From this figure we can see that in
the range of $0$ to $\pi$ the resonant peak becomes sharper as the
phase increases, while in the range of $\pi$ to 2$\pi$ the resonant
peak becomes wider as the phase increases. Fig. 2 is very similar to
Fig. 1(a), which makes clear that, through varying the off-diagonal
term of the self energy, the phase of the BEC reservoir play the
similar role as the Rabi frequency on the transmission probability
$T$.
\begin{figure}
\center
\includegraphics[width=2.5in]{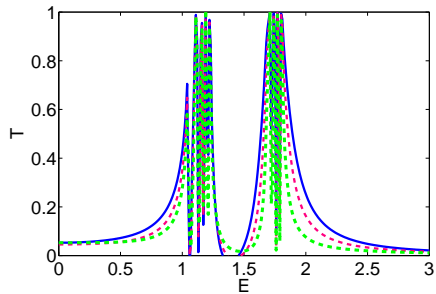}
\caption{(Color Online) The transmission probability $T$ as a
function of the incident boson energy $E$ (in units of $E_{L}$,
$E_{L}$=2.0 kHz), wtih the parameters of
$\delta_{L}$=$\delta_{R}$=0, $\Omega$=0.02 kHz, $g$=10 and
$|\Delta|=2\pi\times$0.41 kHz, the blue solid line for $\phi$=0 or
$2\pi$, the red dashed line for $\phi=\pi/2$ or $3\pi/2$, and the
green dashed line for $\phi=\pi$, respectively.}
\end{figure}

There may be possible experimental realizations for our system.
Firstly, two BEC reservoirs can be realized in current experiments
with atomic gases. Secondly, Luttinger liquid arisen in our systems
can be realized in current experiments with atomic gases. With the
current technology there appears no difficulty in making transverse
frequency $\omega_{\perp}$$>$100$\omega_{z}$, where $\omega_{z}$ is
the longitudinal frequency. In such limit, one can produce atomic
gases with all the atoms lying in the lowest harmonic oscillator
state in the $x$-$y$ plane, leaving the motion along $z$ (the only
degree of freedom). The system then behaves like a 1D Bose gas. For
steeper magnetic traps, $\omega_{\perp}\sim$ 50 kHz, particle
densities of $\rho\sim$ $10^{4}$ particle/cm, and assuming a
scattering length of 110 $a_{B}$ for Rb, it should be possible to
observe the LL behavior \cite{4}.

\section{conclusion}
In conclusion, using the equation of motion for Green function, we
have investigated the transport properties for a Luttinger liquid
coupled to two identical Bose-Einstein condensation reservoirs. It
is demonstrated how the transmission probability is determined by
Rabi frequency, interaction strength, $|\Delta|$, and phase of the
BEC reservoir, respectively. We have found that the distance between
the two resonant transmission probability peaks is determined by the
interaction strengths, while the sharpness of the resonant peak is
mainly determined by the Rabi frequency and phase of the reservoir.
The further theoretical investigation on taking into account
impurity, spin or other interactions are worthy to be carry out.
These results for the proposed system involving a LL may be useful
to control transport properties of cold atom.

{\bf Acknowledgments} We thank Jian-Ming Wang for helpful
discussion. This work was supported by NSF of China under grant
10574042, 90406017, 60525417, 10610335, the NKBRSF of China under
Grant 2005CB724508 and 2006CB921400, Specialized Research Fund for
the Doctoral Program of Higher Education of China (Grant No.
20060542002).

\end{document}